\documentclass[12pt]{article}
\pdfoutput=1 
\usepackage{cite}
\usepackage{a4wide}
\usepackage{graphicx,subfigure}
\usepackage{amsfonts,amstext,amsmath,amssymb,amsthm}
\usepackage[mathscr]{eucal}
\makeatletter

\newcommand{\ii}{{\rm i}}

\begin{document}
\title{\bf Free Fermions with a Localized Source}

\author{P. L. Krapivsky\footnote{Department of Physics,  Boston University, Boston, Massachusetts 02215, USA}
, Kirone Mallick\footnote{ Institut de Physique Th\'eorique, 
CEA Saclay and URA 2306, France}
, Dries Sels\footnote{ Department of Physics,  Harvard  University, Cambridge, Massachusetts 02138, USA}
}
\maketitle

\begin{abstract}
We study an open quantum system of free  fermions on an infinite
lattice coupled to a localized particle source.  In the long time limit,  the total
number of fermions in the system increases linearly with growth rate dependent on the lattice geometry
and dimensionality. We express the growth rate in terms of lattice Green
functions and derive explicit formulae in one dimension and for the square lattice.  The
interplay between the dynamics and the coupling to the environment
leads, in contrast to classical systems, to a non-monotonic dependence of the particle growth rate 
on the input rate. We show that for all lattices the particle growth rate is inversely proportional to the 
input rate when the latter becomes large. This is a manifestation  of the
quantum Zeno effect.
\end{abstract}
\maketitle

\section{Introduction}

Quantum phenomena are  fragile and susceptible to decoherence by  the interplay of an open quantum system with its environment through interactions and measurements.  An environment  that is `constantly watching'  \cite{Haroche} generates  a competition between Hamiltonian dynamics and dissipation \cite{Weiss} that often leads to interesting non-equilibrium  processes. Hence a suitably engineered coupling to a reservoir may be a resource to generate a non-equilibrium state  that can be investigated   experimentally  and analytically \cite{Verstraete,Vengalatorre}.

Classical interacting particle processes  are an ideal laboratory to
explore the differences between thermodynamics and non-equilibrium
phenomena. In this respect,  simple models retaining non-trivial
physical features and amenable to analytical treatment play an
important role in understanding of time-reversal breaking,
rare events, large deviations and dynamical phase transitions
\cite{SpohnBook,D07,Bertini}. Coupling  to reservoirs represented by
Markov processes  have found numerous applications to non-equilibrium
dynamics  ranging from biophysical models to urban traffic flows
\cite{PaulBook}. One such  paradigmatic  model is an exclusion
process---a lattice gas  of  random walkers subject to a  hard-core
constraint that prevents two walkers from being simultaneously at the
same site \cite{Chou}.  This  {\it classical} exclusion  process has
been instrumental to exploring non-equilibrium states generated by
various means such as unbalanced reservoirs, internal driving field or
relaxation from a specific initial  condition. For example, the study
of the exclusion process on an  infinite   lattice  has led
to a better understanding of current statistics, height distribution
in the Kardar-Parisi-Zhang equation and their  relations  with
Tracy-Widom distributions from random matrix theory (see
\cite{Takeuchi,Krug,Gershenfeld} and references therein).  Another
example is the symmetric  exclusion process driven by a localized
source \cite{PaulSEP,DarkoSEP}. The statistics of the total number
$\mathcal{N}(t)$ of particles entering an initially empty system
displays a strong dependence on dimension \cite{PaulSEP}. For
instance, the average number of particles
$N(t)=\langle\mathcal{N}(t)\rangle$  grows according to\footnote{We
  tacitly assume that the underlying lattice is hyper-cubic if not
  stated otherwise, so the Watson integral \eqref{Watson-int}
  corresponds to the lattice ${\mathbb Z}^d$ with $d\geq 3$.  Equation
  \eqref{Nav:SEP} accounts for simple exclusion processes with total
  hopping rate equal to 2 for all $d$.} 
\begin{equation}
\label{Nav:SEP}
N(t) \simeq 
     \begin{cases}
  4 \sqrt{ \frac{t}{\pi}}  \quad    &d =1 \\
  \frac{2 \pi t}{\ln t} \quad  \quad   &d =2 \\
  \frac{\Gamma}{1 + \Gamma W_d}\, t   \quad  \quad   &d >2  
\end{cases} 
\end{equation}
where $\Gamma$ is the injection rate and $W_d$  the  Watson integral defined via
\begin{equation}
\label{Watson-int}
W_d = \frac{d}{2} \int_{0}^{2 \pi} \ldots
\int_{0}^{2 \pi} \prod_{i=1}^d \frac{d q_i}{2 \pi} \, 
\frac{1}{ \sum_{i=1}^d \left(1 - \cos q_i \right)} 
\end{equation}
Interestingly, in low dimensions, viz. for $d=1$ and $d=2$, the leading behavior in \eqref{Nav:SEP} is independent of  the injection rate.

In the present work we study a similar situation in the quantum setting, namely we analyze an open quantum system of fermions driven by a source  that injects fermions at the origin. We consider the simplest non-interacting (spinless) fermions; the Pauli exclusion makes this quantum system somewhat analogous to the exclusion process.\footnote{This analogy is intuitively appealing, but somewhat misleading:  the one-dimensional  symmetric exclusion process can be mapped exactly to  the Heisenberg XXX spin-chain in imaginary time and not to free fermions (see also \cite{Eisler11}).}  Our goal  is to  study how the number $N$ of fermions in the system increases with time, as a function of the injection rate  $\Gamma$ and the dimensionality $d$ of the lattice.\footnote{A closely related quantum process defined in continuous space was analyzed in \cite{Spohn10}.}  The growth of the number of fermions differs significantly from the corresponding behavior of the classical exclusion process \eqref{Nav:SEP}. Some qualitative behaviors are simpler and more intuitive in the quantum case. In particular, the average number of fermions increases linearly with time, $N\simeq C_d(\Gamma) t$, regardless of the spatial dimension $d$. This is easy to appreciate---quantum walks are ballistic in any dimension \cite{Kempe}, and so the effective jamming of the lattice in the vicinity of the origin by classical particles that  leads to the sub-linear growth when $d\leq 2$, see \eqref{Nav:SEP}, is avoided.  The injection rate affects the asymptotic behavior in all $d$ and the particle growth rate has an expected asymptotic $C_d(\Gamma)\sim \Gamma$ when $\Gamma\to 0$. A counter-intuitive behavior occurs in the opposite limit, $\Gamma \to \infty$, where $C_d(\Gamma)\sim 1/\Gamma$. This is a signature of the quantum Zeno effect.
  
The outline of this work is as follows. In the next section ~\ref{sec:Model}, we define the model using the Lindblad equation formalism and present a few chief results. In section~\ref{sec:1d},  we calculate the growth rate and the density profile for the one-dimensional lattice.  In section~\ref{sec:high-d}, a  general approach valid for an  arbitrary dimension is presented  and explicit calculations are performed for the square lattice. Various extensions are discussed in section \ref{disc}. Details of the calculations and an overview of the lattice Green functions are presented in the appendices.

\section{The model and some exact  results}
\label{sec:Model}

We  study an open quantum system on an infinite lattice driven by a localized source of fermions. The system is  initially empty;   at time $t=0$,  the  source is turned on and fermions are  injected at the origin at a  constant rate. 

The evolution of open quantum systems is described by the Lindblad equation \cite{Sud76,Lindblad76,Breuer,Preskill,Manzano}
\begin{equation}
\label{Lindblad}
\partial_t \rho = -\ii [H, \rho] + 2L \rho L^{\dagger}-\{L^{\dagger}L,
\rho\}
\end{equation}
for the density matrix $\rho(t)$. The set of operators $L$
describe the interactions with reservoirs. In our situation when the
fermions are injected at the single site which we set to be the origin, we have a
single operator $L=\sqrt{\Gamma}\, c_{\bf 0}^{\dagger}$ modeling the
insertion with intensity $\Gamma$. The curly brackets in
equation~\eqref{Lindblad} denote the anti-commutator, $\ii = \sqrt{-1}$ and
we set $\hbar=1$. We consider identical non-interacting spinless lattice fermions described by the Hamiltonian
$H =  \sum_{\langle {\bf i}, {\bf j} \rangle}( c_{\bf i}^{\dagger}  c_{\bf j} + c_{\bf j}^{\dagger}  c_{\bf i})$,
where the sum is taken over pairs of neighboring lattice sites ${\bf i}$ and ${\bf j}$.
We consider uniform systems, so the hopping rates are equal and set to unity.
Fermions satisfy the Pauli exclusion principle:  $\{ c_{{\bf i}}^{\dagger} ,  c_{{\bf j}} \} = \delta_{ij}$
and all the  other anticommutators vanish. In one dimension, the Hamiltonian reads 
\begin{equation}
\label{Ham}
H =  \sum_{n=-\infty}^\infty \left( c_n^{\dagger}  c_{n+1} +
c_{n+1}^{\dagger} c_{n} \right)
\end{equation} 

A basic characteristic of this  open quantum system is the total number of particles $\mathcal{N}(t)$ at time $t$. The
full statistics of this evolving random quantity is described by the
probability distribution $P(N,t)=\text{Prob}[\mathcal{N}(t)=N]$. We
focus on the average $N(t) = \langle \mathcal{N}(t)\rangle$ and show
that it exhibits a linear growth
\begin{equation}
\label{N-linear}
N(t)\simeq C_d(\Gamma)\,t
\end{equation}
when $t\gg 1$. The particle growth rate $C_d(\Gamma)$  depends on the strength
 $\Gamma$ of the source and on the spatial dimension $d$
(and also on the geometry of the underlying lattice).

It is possible to obtain a formal expression for the growth rate in any dimension in terms of lattice Green
functions (see Appendix~\ref{app:LGF}), and we succeeded in deriving explicit tractable formulae
for $d=1$ and $d=2$ (which are experimentally most relevant). We now state  some of the results  derived in the following sections.

In one dimension (section~\ref{sec:1d}), the fermion growth rate $C_1(\Gamma)$ is given by 
\begin{equation}
\label{1d-main}
C_1(\Gamma) =  2\Gamma -
\frac{2\Gamma^2}{\pi}\left[\gamma^{-2}+\left(\gamma^{-1}-\gamma^{-3}\right)\tan^{-1}
  \gamma \right]
\end{equation}
with 
\begin{equation}
\label{gamma}
\gamma=\sqrt{(\Gamma/2)^2-1}
\end{equation}
Note that equation~\eqref{1d-main} 
is formally valid when $\Gamma>2$ and  $\gamma$
is real and positive. For $\Gamma<2$, the result remains  
 valid after analytical continuation. 

From equation~\eqref{1d-main},  one can derive extremal behaviors. When $\Gamma\ll 1$, we have
\begin{subequations}
\begin{equation}
\label{C1:small}
C_1(\Gamma) = 2\Gamma +  \frac{4\Gamma^2}{\pi}\left[\ln
  \frac{\Gamma}{4} + \frac{1}{2}\right]+ O(\Gamma^4\ln \Gamma)
\end{equation}
For large insertion rate, $\Gamma\gg 1$, the growth rate decays as  
\begin{equation}
C_1(\Gamma) = 4\Gamma^{-1} - \tfrac{128}{3 \pi}\,\Gamma^{-2}+36
\Gamma^{-3} - \tfrac{4096}{15 \pi}\,\Gamma^{-4}+\ldots
\end{equation}
\end{subequations}

\begin{figure}
\centering 
\subfigure[]{\includegraphics[width=0.44\textwidth]{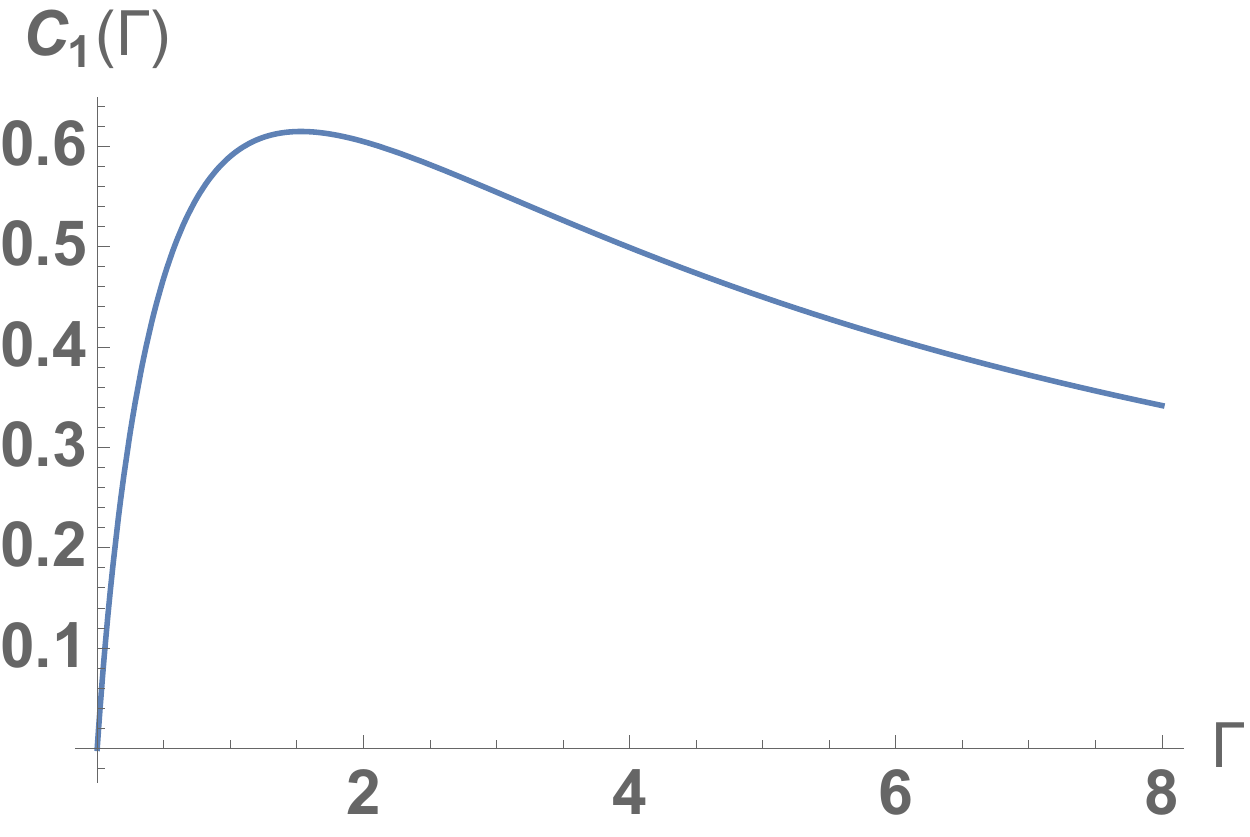}}\qquad
\subfigure[]{\includegraphics[width=0.44\textwidth]{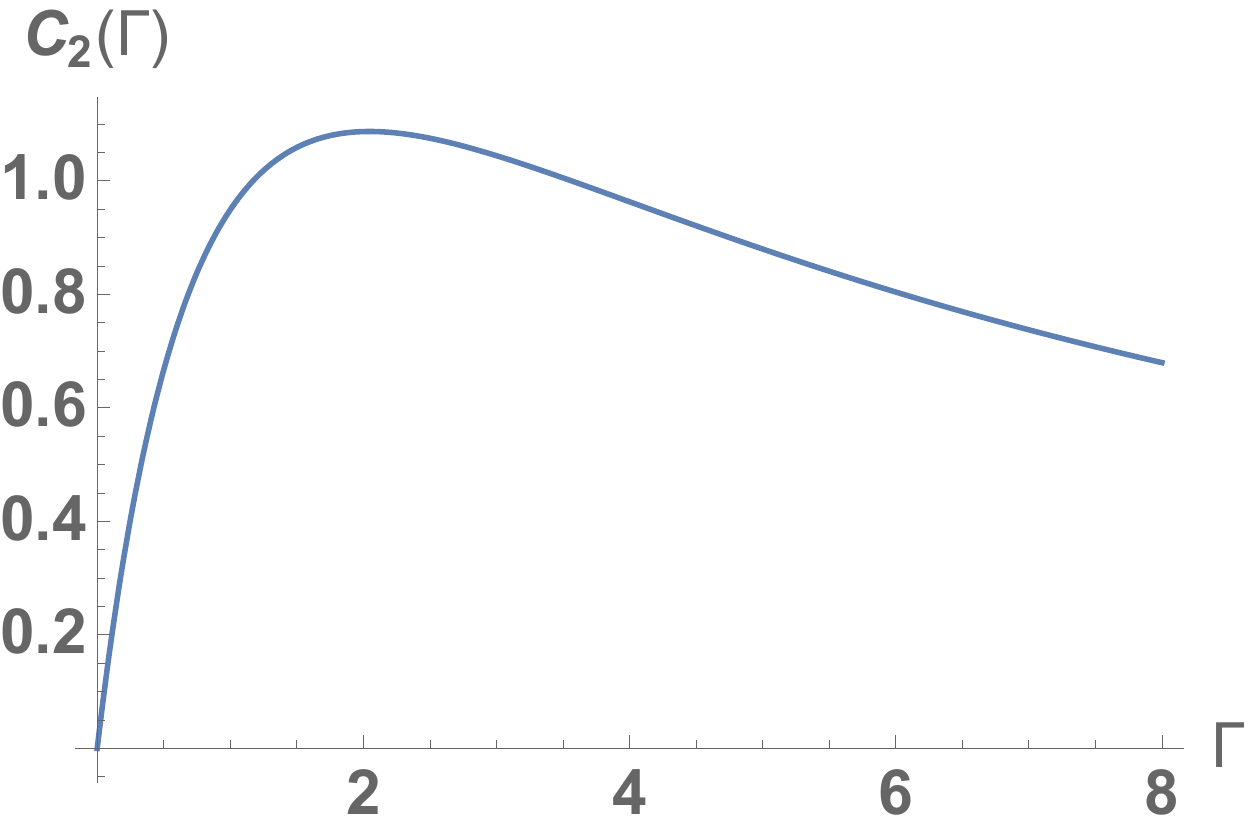}}
\caption{(a) Particle growth rate on the  one-dimensional lattice:
   $C_1(\Gamma)$ is defined via \eqref{1d-main}. The maximum $C_1^\text{max}\approx
  0.614798$ is reached at $\Gamma_1\approx 1.5313$.
  (b) Growth rate for  the square lattice. The analytical
  prediction given by \eqref{2d-main} and \eqref{I1}--\eqref{I2} is
  plotted. The maximum value $C_2^\text{max}\approx 1.08984$ is
  reached at $\Gamma_2\approx 2.04602$.}
\label{Fig:C1-2-exact}
\end{figure}

The particle growth rate $C_1(\Gamma)$ increases with $\Gamma$ for
sufficiently small insertion rates and then decreases asymptotically
approaching to zero when $\Gamma\to\infty$, see Fig.~\ref{Fig:C1-2-exact} (a).  
The vanishing of the particle growth rate when $\Gamma\to\infty$ is the manifestation of the quantum Zeno
effect (QZE),  one of the simplest and most paradigmatic differences between
classical and quantum  open  systems. The QZE
 refers to the counter-intuitive phenomenon that a
frequent measurement of a quantum system does not allow the state to
evolve unitarily \cite{Q_Zeno,QZE:PRA,QZE:entang}.
For weak measurements, the system is only slightly perturbed 
by the measurement back-action. When the measurements are
strong, the system gets projected to an eigenstate of the
measurement operator (the wavefunction collapses) and if one keeps on 
measuring, the system is unable to evolve. Fundamentally, there is little
difference between continuous measurements and a dissipative process
as the former entails connecting the system to a large reservoir (the
measurement device). The only real difference is that measurements are
typically recorded. It is thus quite natural that 
the dynamics of simple lattice problems with local injection or removal 
of particles exhibits the QZE.

The QZE has been demonstrated in various experiments with ultra-cold
atoms, see e.g. \cite{Cirac08,Zoller09,Cirac10}. Some recent experiments
\cite{Ott13,Ott16,Ott18} are particularly close to our model. The
experimental setup \cite{Ott13,Ott16,Ott18} involves a local loss
process induced by shining a focused electron beam onto an atomic
Bose-Einstein condensate  residing on the one-dimensional optical
lattice. The QZE manifests itself in the number of atoms lost from the
condensate:  linear growth was found for small $\Gamma$  and an 
inverse $1/\Gamma$ scaling was observed when $\Gamma$ is large. A
recent theoretical analysis \cite{Kollath} of this one-dimensional
system with local losses  combines the Lindblad framework with
continuum modeling based on the Luttinger liquid description and
emphasizes the similarity with the Kane-Fisher barrier problem
\cite{KF92a,KF92b}.

In higher dimensions,  exact results for the Laplace
transforms are given in Section~\ref{sec:high-d}. For the square lattice
\begin{equation}
\label{2d-main}
C_2(\Gamma) =  2\Gamma -
\tfrac{16}{\pi}\,\Gamma^2\big[I_1(\Gamma)+I_2(\Gamma)\big]
\end{equation}
where $I_1$ and  $I_2$ are the integrals  
\begin{subequations}
\begin{align}
\label{I1}
I_1(\Gamma) &= \int_0^1 dx\,\frac{K^2(x)+K^2(x')}{[\Gamma
    K(x)]^2+[\Gamma\ K(x')+2\pi]^2}\\
\label{I2}
I_2(\Gamma) &= \int_0^1 \frac{dx}{\Gamma^2 x^2+[2\pi/K(x)]^2}
\end{align}
\end{subequations}
and  $K(x)$  is  the complete elliptic integral of the first kind of modulus $x$:
\begin{equation}
\label{Kx:def}
K(x) = \int_0^{\pi/2}\frac{d\theta}{\sqrt{1-x^2\sin^2\theta}}
\end{equation}
In \eqref{I1} we have  used the standard notation $x'=\sqrt{1-x^2}$ to denote the 
complementary modulus.  

The dependence of the particle growth rate $C_2(\Gamma)$ on the
insertion rate $\Gamma$ is qualitatively the same as in one dimension,
see  Fig.~\ref{Fig:C1-2-exact} (b).  When $\Gamma\ll 1$, the particle growth rate can be written as
\begin{equation}
\label{small-2d}
C_2(\Gamma) = 2\Gamma -A_2\Gamma^2+A_3\Gamma^3+\ldots
\end{equation}
with 
\begin{equation}
\label{small-2d-A23}
\begin{split}
A_2 &= \frac{4}{\pi^3} \int_0^1 dx\left[2K^2(x)+K^2(x')\right] =
1.8054571566\ldots \\ A_3 &= \frac{4}{\pi^3} \int_0^1
dx\left[K^2(x)\,K(x')+K^3(x')\right] =  1.29401145979  \ldots
\end{split}
\end{equation}
The asymptotic expansion \eqref{small-2d} is obtained  by expanding \eqref{I1}--\eqref{I2} when
$\Gamma\ll 1$ and it is in excellent agreement with the exact prediction for
sufficiently small $\Gamma$. 
When $\Gamma \to \infty$, the  leading behavior is given by 
\begin{equation}
\label{large-2d}
C_2(\Gamma) \simeq \frac{8}{\Gamma} 
\end{equation}
The universal decay, $C_d\simeq B_d\Gamma^{-1}$ as $\Gamma\gg 1$, occurs in
all spatial dimensions and for hyper-cubic lattices the amplitude  has a  simple value 
\begin{equation}
\label{Cd-large}
\lim_{\Gamma\to\infty} \Gamma C_d(\Gamma) = B_d = 4d
\end{equation}
The universal $\Gamma^{-1}$ decay in all dimensions is a signature of the QZE,
which  is a local dynamical effect, due to the measurement, and is  not related to the global
geometry of the system  under study. Note that the  dimension $d=2$ which separates
transient and recurrent behavior for classical random walks plays no special role
 in the quantum case.

\section{Free fermions on $\mathbb{Z}$ with a source}
\label{sec:1d}

The basic quantities that we wish to determine are 
the average total number of particles  as a function of time
and the particle density profile. Both observables 
are encoded in the two-point correlation functions, defined as 
\begin{equation}
\label{sigma:def}
 \sigma_{i,j}(t) = \langle c_i^{\dagger} c_j  \rangle_t = {\rm Tr}\left(\rho(t) c_i^{\dagger} c_j \right)
\end{equation}
Using the Lindblad equation \eqref{Lindblad} with Hamiltonian \eqref{Ham} and $L=\sqrt{\Gamma}\, c_0^{\dagger}$ we deduce that 
the  two-point correlations \eqref{sigma:def} satisfy a closed system\footnote{A similar property holds for the symmetric exclusion process \cite{Gershenfeld} and for  some of its quantum generalizations \cite{Eisler11}.}  of differential equations
\begin{eqnarray}
\label{sigma}
 \frac{d  \sigma_{i,j}}{dt} &=& \ii \left( \sigma_{i+1,j} +
 \sigma_{i-1,j} - \sigma_{i,j+1} -  \sigma_{i,j-1} \right)
 - \Gamma\left(\delta_{i,0} \sigma_{i,j}+ \delta_{j,0}
 \sigma_{i,j}\right) + 2\Gamma\delta_{i,0} \delta_{j,0}
\end{eqnarray}
These equations are linear and exactly solvable. 

It is more convenient to deal with equations
\begin{eqnarray}
\label{sigma-dual}
 \frac{d  \sigma_{i,j}}{dt} &=& \ii \left( \sigma_{i+1,j} +
 \sigma_{i-1,j} - \sigma_{i,j+1} -  \sigma_{i,j-1} \right)
 -  \Gamma\left(\delta_{i,0} \sigma_{i,j}+ \delta_{j,0}
 \sigma_{i,j}\right)
\end{eqnarray}
which are identical to \eqref{sigma} apart from the source term, the last term on the right-hand side of \eqref{sigma}, which is absent in
\eqref{sigma-dual}. Interestingly, \eqref{sigma-dual} describe the {\it dual problem} in which particles are {\it removed} from
 the origin at rate $\Gamma$; this process is modeled by  the Lindblad equation \eqref{Lindblad} with the same Hamiltonian \eqref{Ham} and $L=\sqrt{\Gamma}\,  c_0$. In order to calculate the growth rate we shall analyze this dual problem starting with a single particle at the origin. 
 
 A crucial observation simplifying the analysis of \eqref{sigma-dual} is that the solution of \eqref{sigma-dual} factorizes, that is  the ansatz
\begin{equation}
  \sigma_{i,j}(t) = S_i(t) S_j^*(t)
  \label{2pt-ansatz}
\end{equation}
is consistent with \eqref{sigma-dual} if the functions  $S_n(t)$ satisfy 
\begin{equation}
\label{S-dual}
 \frac{d  S_n}{dt} = \ii [S_{n+1} +  S_{n-1}] - \Gamma \delta_{n,0}
 S_n
\end{equation}
with  initial condition 
\begin{equation}
\label{IC}
S_n(0) = \delta_{n,0}
\end{equation}
The initial-value problem \eqref{S-dual}--\eqref{IC} represents
a continuous-time quantum walk, namely a quantum particle
propagating on a  chain, in the presence of a trap
of strength $\Gamma$ at the origin.  The survival of a quantum
particle subject to a complex optical  potential
\cite{Pearlstein,Parris,Selsto1,Selsto2} was  analyzed 
in  \cite{KLM14}. The factorization~\eqref{2pt-ansatz} shows that, in the present case,
this non-Hermitian potential is equivalent to the more
rigorous framework provided by the Lindblad equation 
that also includes the measurement process
(see \cite{Dhar1,Dhar2,Dhar3,Barkai1,Barkai2,Barkai3} for related
 approaches in the context of first detection time of a quantum walk).

\subsection{Solution of the one-body problem}

Equation \eqref{S-dual} has been analyzed in the past, see \cite{KLM14} and references therein. The approach is based on the Laplace-Fourier transform. Performing the Laplace transform with respect to time,
\begin{equation}
\widehat{S}_n(s) = \int_0^\infty dt\,e^{-st} S_n(t) \, , 
\end{equation}
we recast \eqref{S-dual} into
\begin{equation}
\label{S-Lap}
s\widehat{S}_n - \delta_{n,0} = \ii \big[\widehat{S}_{n+1} +
  \widehat{S}_{n-1}\big] - \Gamma \delta_{n,0} \widehat{S}_n
\end{equation}
Performing the Fourier transform in space, 
\begin{equation}
\label{S-Fourier}
S(s,q) =  \sum_{n=-\infty}^\infty \widehat{S}_n(s)\,e^{- \ii qn}\,, 
\end{equation}
we rewrite equation  \eqref{S-Lap} as
\begin{equation}
sS-1 = 2 \ii S \cos q -  \Gamma
\widehat{S}_0
\end{equation}
yielding 
\begin{equation}
\label{SS0}
S = \left[1 -  \Gamma \widehat{S}_0(s)\right] \Psi(s, q), \quad
\Psi(s, q)=\frac{1}{s - 2 \ii \cos q}
\end{equation}
The definition \eqref{S-Fourier} implies 
\begin{equation*}
\widehat{S}_0(s) = \frac{1}{2\pi} \int_0^{2\pi} dq\,S(s,q)
\end{equation*}
which in conjunction with  \eqref{SS0} fixes 
\begin{equation}
\label{S0}
\widehat{S}_0(s) = \frac{1}{\Gamma+\sqrt{s^2+4}}
\end{equation}
and
\begin{equation}
\label{S-sol}
S(s,q) = \frac{\sqrt{s^2+4}}{\Gamma+\sqrt{s^2+4}} \,\Psi(s,q)
=  \frac{1}{\Gamma+\sqrt{s^2+4} } \,
 \frac{ \sqrt{s^2+4}  } {s - 2 \ii \cos q}
\end{equation}
Inverting the Fourier transform, we deduce 
\begin{equation}
\label{Sn-Lap}
\widehat{S}_n(s) = \frac{1}{\Gamma+\sqrt{s^2+4}}
\left(\frac{2\ii}{s+\sqrt{s^2+4}}\right)^{|n|}
\end{equation}
The inverse Laplace transform of this expression
determines the  fermion density $S_n(t)$ at a given time
and at position $n$ (see  Appendix~\ref{App:density-dual-1d-singlep}).

 \subsection{Calculation of the growth rate}

We now calculate the growth rate $C_1(\Gamma)$.
The average total number of particles can be expressed in terms of 
 the correlation functions
\begin{equation}
\label{N:sigma}
N(t) =  \sum_{n=-\infty}^\infty \sigma_{n,n}(t) 
\end{equation}
where the $\sigma_{i,j}(t)$'s satisfy the differential equations
\eqref{sigma} with initial condition  $\sigma_{i,j}(0) = 0$ for all $i,j$.
Note that the linear system \eqref{sigma}  can formally be rewritten as
\begin{equation}
  \dot \sigma = M  \sigma + V
  \label{sigma:formel}
  \end{equation}
Here  $\sigma=\sigma(t)$ is an infinite vector with components $\sigma_{i,j}(t)$;
$M$ is the  matrix that encodes the homogeneous terms\footnote{The same $M$ arises in the dual problem of removing a particle
from the origin.}; and $V$,  the source term  in \eqref{sigma}, is a vector with a unique non-zero component, 
 $2 \Gamma \delta_{i,0} \delta_{j,0}$. Equation \eqref{sigma:formel}
is solved to yield 
\begin{equation}
   \sigma(t) = \int_0^t {\rm d}\tau \, {\rm e}^{(t - \tau) M} V
\end{equation}
The vector ${\rm e}^{(t - \tau) M} V$ represents two-point correlations in  the dual problem  \eqref{sigma-dual}
with a fermion starting at the origin. Hence, using the ansatz \eqref{2pt-ansatz}, we deduce
\begin{equation}
\label{sigma:S}
  \sigma_{i,j}(t)  =   2 \Gamma \int_0^t {\rm d}\tau  S_i(\tau) S_j^*(\tau)
\end{equation}
where the functions $S_i$ satisfy  the dynamics \eqref{S-dual}. Combining \eqref{N:sigma} and \eqref{sigma:S} we obtain
\begin{equation}
  N(t) = 2 \Gamma \int_0^t {\rm d}\tau \, \sum_n | S_n(\tau)|^2
         = 2\Gamma \int_0^t  {\rm d}\tau \, \Pi(\tau)
 \end{equation}
where $\Pi(t)$ represents the survival probability of a single fermion starting at the origin, that
can be removed at the origin with rate $\Gamma$. This survival probability
\begin{equation}
\Pi(t) = \sum_{n=-\infty}^\infty |S_n(t)|^2= 1- 2\Gamma \int_0^t dt'\,
|S_0(t')|^2
\end{equation}
approaches,  when $t \to \infty$,   to
\begin{equation}
\label{final}
\Pi_\infty =  1- 2\Gamma I, \quad I =  \int_0^\infty dt\, |S_0(t)|^2
\end{equation}
Using  the Parseval-Plancherel identity
\begin{equation}
\label{PP}
\int_0^\infty dt\,f(t) g(t) = \int_{-\ii \infty }^{\ii
  \infty}\frac{ds}{2\pi \ii}\, \widehat{f}(s) \widehat{g}(-s)
\end{equation}
for the Fourier transforms and setting $f(t) = S_0(t)$ and $g(t) =
S_0^*(t)$,
we express the integral in \eqref{final}  through the Laplace
transform $\widehat{S}_0(s)$.  This approach yields
\cite{KLM14}
\begin{equation}
\label{int}
 I = \frac{1}{\pi}\int_0^2 \frac{dy}{\big(\Gamma+\sqrt{4-y^2}\big)^2}
 +  \frac{1}{\pi}\int_2^\infty \frac{dy}{\Gamma^2+y^2-4}
\end{equation}
When $\Gamma>2$, we compute the integrals and arrive at 
\begin{equation}
\label{>2}
\Pi_\infty =  1-
\frac{\Gamma}{\pi}\left[\gamma^{-2}+\left(\gamma^{-1}-\gamma^{-3}\right)\tan^{-1}
  \gamma\right]
\end{equation}
with $\gamma\equiv \sqrt{(\Gamma/2)^2-1}$; for $\Gamma<2$, the
survival probability can then be obtained by analytical continuation of \eqref{>2}. 

Thus, we conclude that the average number of fermions grows linearly in time and  
\begin{equation}
\label{ampl}
C_1(\Gamma) = 2\Gamma \Pi_\infty
\end{equation}
leading to  the announced result
\eqref{1d-main} for the particle growth rate.

\subsection{The density profile}
\label{sec:density}

We wish to  determine the density profile of the particles at time $t$ starting with an initially empty  system. It is equivalent to studying  the dual problem of a system initially full of holes, where holes are removed at the origin (and replaced by particles).
In this formulation, the governing equations
are given by  \eqref{sigma-dual} and the initial conditions become
\begin{equation}
\label{IC-holes}
 \sigma_{i,j}(0) = \delta_{i,j}
\end{equation}
The answer to  this  problem can be written as a linear
combination
\begin{equation}
\sigma_{i,j}(t) = \sum_k S_i^{(k)}(t)\left( S_j^{(k)}(t) \right)^*
\end{equation}
where  $S_n^{(k)}(t)$ solves
the one-body non-Hermitian dynamics \eqref{S-dual}
with initial condition
\begin{equation}
  S_n^{(k)}(0) =  \delta_{n,k}
\end{equation}
This one-body problem is solved by the Laplace transform
and we obtain
\begin{eqnarray}
  \widehat{S_n^{(k)}}(s) & = & \frac{1}{\sqrt{s^2+4}}
  \left(\frac{2\ii}{s+\sqrt{s^2+4}}\right)^{|n-k|} \nonumber\\
  &-&  \frac{\Gamma}{\Gamma+\sqrt{s^2+4}}\frac{1}{\sqrt{s^2+4}}
  \left(\frac{2\ii}{s+\sqrt{s^2+4}}\right)^{|n|+|k|}
  \label{sol:dual-laplace}
\end{eqnarray}
Performing the inverse Laplace transform yields  
\begin{equation}
  S_n^{(k)}(t) = i^{|n-k|} J_{|n-k|}(2t) - \Gamma {\ii}^{|n| + |k|}
   G_{|n| + |k|}(t)
  \label{sol-pour-Sn}
\end{equation}
where  $J_{|n-k|}(2t)$ is a Bessel function and $ \Gamma {\ii}^{|n| + |k|} G_{|n| + |k|}(t) $ denotes  the inverse Laplace transform of
the expression in the second line on  the right-hand side of Eq.~\eqref{sol:dual-laplace}.

We conclude that  the density of particles  at site $n$ is given by
\begin{equation}
  N_n(t) = 1 - \sigma_{n,n}(t) = 1 -  \sum_k | S_n^{(k)}(t)|^2
\label{eq:profil}
\end{equation}
with $S_n^{(k)}(t)$ given by \eqref{sol-pour-Sn}. To extract the long time behavior, we re-parametrize $k$ and $n$:
\begin{equation}
 k = 2t u, \qquad n =  2t v
\end{equation}
Due to symmetry, it suffices to consider the $n\geq 0$ range, so $v \geq 0$. In the $ t \to \infty$ limit,  the density profile is obtained by applying the saddle-point method for the functions $J_{n}$ and  $G_{m}$. The asymptotic formulae read
\begin{eqnarray}
 \label{asymptJn} 
   J_{|n-k|}(2t) & \simeq & 
   \frac{ \cos \big[2t \sqrt{(1 - (u -v)^2} - 2 t |u-v|  \arccos |u-v| -\frac{\pi}{4} \big]}
   {\sqrt{\pi t}\, \left[1 - (u -v)^2\right]^{1/4}} \\
 \label{asymptGn}     
   G_{|n| + |k|}(t)& \simeq &
   \frac{ \cos \big[2t \sqrt{(1 - (v +|u|)^2} - 2  t (v +|u|) \arccos (v +|u|)
      -\frac{\pi}{4} \big]}
        {\sqrt{\pi t}\, \left[\Gamma + 2(v + |u|)\right]\left[1 - (v + |u|)^2\right]^{1/4}}
 \end{eqnarray}
Equation \eqref{asymptJn}  is standard, see \cite{Abram}; equation \eqref{asymptGn} is derived in a similar fashion. The asymptotic formulae \eqref{asymptJn}--\eqref{asymptGn} are valid in the region $ - 1 \le u - v \le 1$ and  $ - 1 \leq |u| +  v \leq 1$. Recalling that $v \geq 0$ one gets $ |u| \le 1 -v$. 

In the long  time limit, the profile is obtained by substituting (\ref{asymptJn}) and (\ref{asymptGn}) into (\ref{sol-pour-Sn}) and (\ref{eq:profil}). After converting the discrete sum over the integer $k$ into integrals with respect to the continuous variable $v$  we find   the density at the position $n$ to be 
\begin{equation}
 \label{NwI}
   N_n(t) = {\mathcal N}(v) + I_n 
\end{equation}
The profile ${\mathcal N}(v)$ describes the density at the bulk where $v  = \frac{n}{2t}>0$ and it has the form
\begin{equation}
 \label{Nw}
   {\mathcal N}(v)  = \frac{2 \Gamma}{\pi} \int_v^1
   \frac{ {\rm d}x}{\sqrt{1 -x^2}} \frac{ 2 x} {\left( \Gamma + 2 x\right)^2}
\end{equation}
Near the origin there is a spike, namely an inner layer of width $n=O(1)$ where the density profile deviates from the continuous density profile ${\mathcal N}(v) $ that varies on the scale $n=O(t)$.  The deviation is given by
\begin{equation}
 \label{In}
   I_n =   \frac{2 \Gamma}{\pi} \int_0^{\frac{\pi}{2}}
  \frac{ \cos( 2 n \theta + n \pi)} { \Gamma + 2 \cos \theta}\, {\rm d}\theta  
\end{equation}
This correction  decays rapidly  when $|n|\gg 1$. Using integration by parts one extracts the asymptotic approach of the spike near the origin to  the bulk density:
\begin{equation}
 \label{In:asymp}
I_n=\frac{1}{\pi \Gamma}\,\frac{1}{n^2}+O(n^{-4})
\end{equation}

\begin{figure}[t]
	\centering
\subfigure[]{\includegraphics[width=0.45\textwidth]{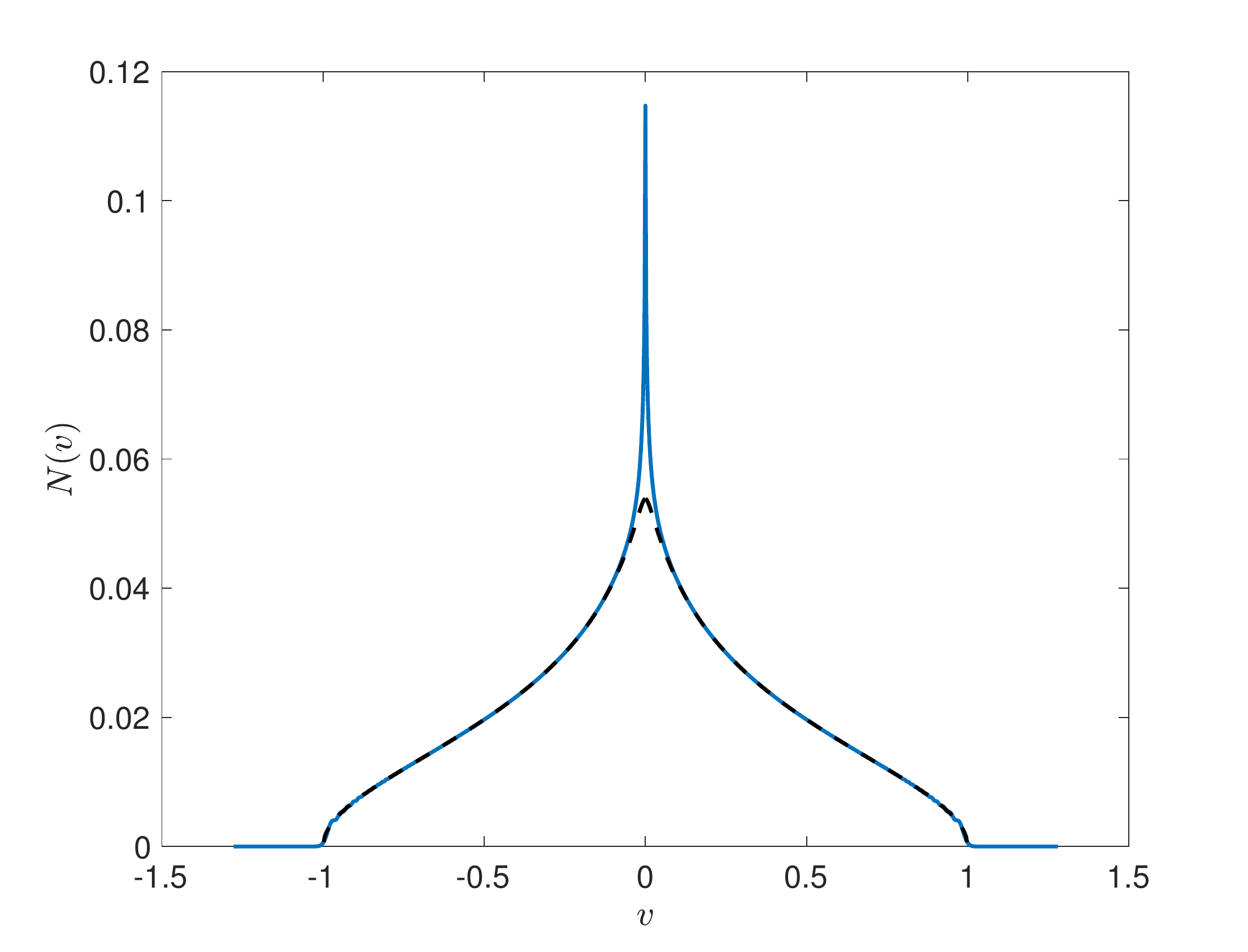}}\qquad
\subfigure[]{\includegraphics[width=0.45\textwidth]{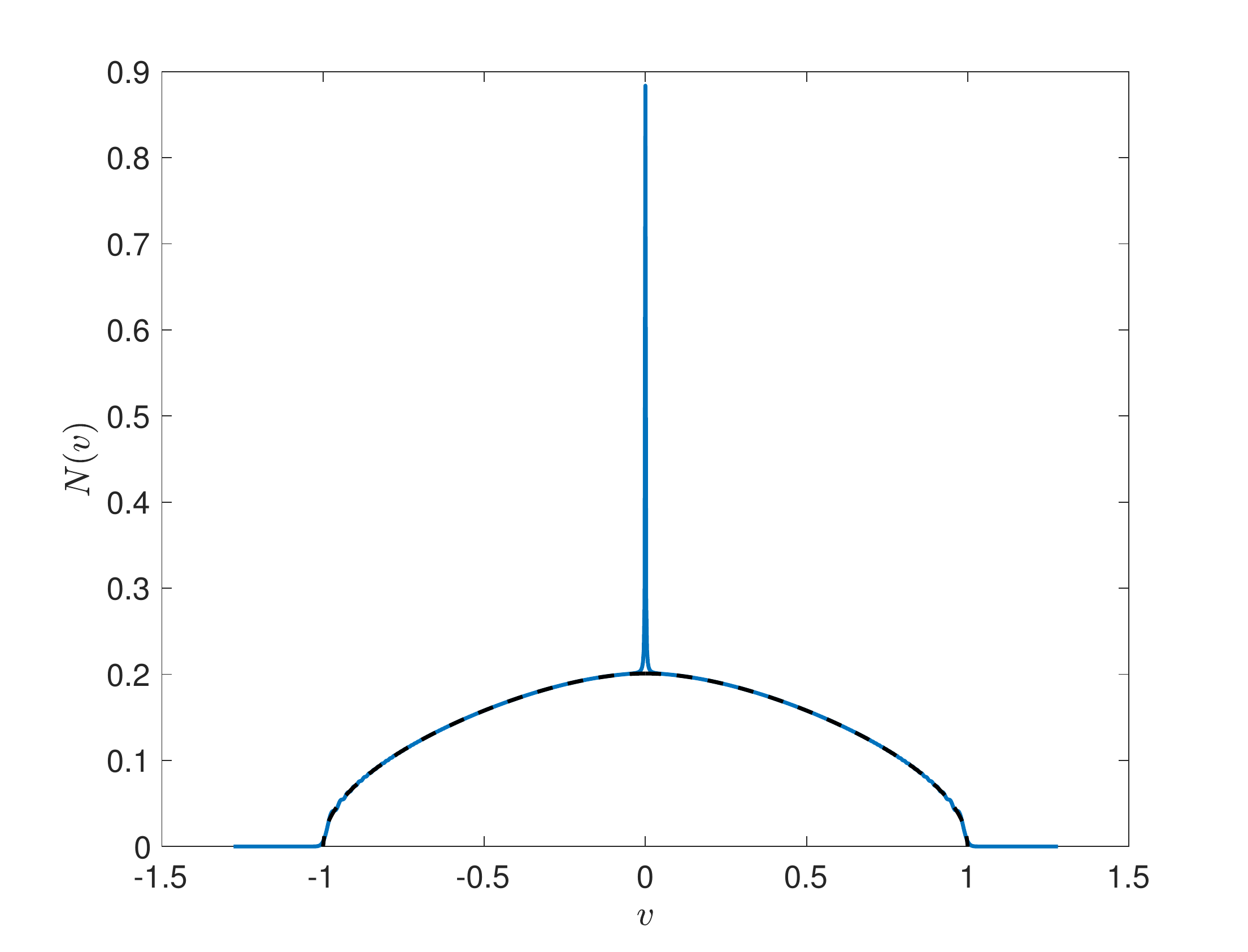}}
	\caption{Density profile of injected particles for (a) $\Gamma=0.05$ and (b) $\Gamma=2.5$. Blue line shows the numerical exact result obtained at a time $t=200$ in rescaled coordinates and the dashed line shows $\mathcal{N}(v)$. }
	\label{fig:density}
\end{figure}

Figure~\ref{fig:density} shows a comparison of the asymptotic bulk density profile $\mathcal{N}(v)$ to exact numerical results at finite time. Note the excellent agreement in the bulk. The spike near the origin described by $N_n(t)\simeq {\mathcal N}(0) + I_n$ with $I_n$ given by \eqref{In} is also in good agreement with numerical results for finite time. There are also oscillations near the causal horizon caused by the dispersion of the particles. These oscillations are not accounted by our asymptotic formulas, but they can be deduced from the exact results \eqref{sol:dual-laplace}--\eqref{eq:profil} using a more accurate treatment in the region $|n|-2t = O(t^{1/3})$. This $t^{1/3}$ scaling of the width of the front region and its staircase like internal structure arises in many problems involving free fermions, see e.g. \cite{Zoltan1,Zoltan2}.

Setting $n=0$ in equations \eqref{NwI}--\eqref{In},  we find  the density at the origin:
\begin{eqnarray}
N_0 = 1 -  \frac{2}{\pi} \int_{0}^1 \frac{ {\rm d}x}{\sqrt{1 -x^2}}\left(  \frac{ 2 x} {\Gamma + 2 x}\right)^2 
\end{eqnarray}
Computing the integral gives 
\begin{equation}
\label{N0-more}
N_0 = \frac{\Gamma}{\pi}\left[\gamma^{-2}+\left(\gamma^{-1}-\gamma^{-3}\right)\tan^{-1} \gamma\right]
\end{equation}
with $\gamma = \sqrt{(\Gamma/2)^2-1}$ when $\Gamma>2$;  when $\Gamma<2$, the result follows from \eqref{N0-more} by analytical continuation. As expected, $1 - N_0 = \Pi_\infty$ with $\Pi_\infty$ given in \eqref{>2}. As another consistency check we compute the total average number of fermions 
\begin{eqnarray*}
   \sum_{n \in {\mathbb Z}}  N_n(t) & =  &N_0(t) + 2 \sum_{n=1}^\infty N_n(t) \nonumber \\
&\simeq &  2 \times 2 t \,\frac{2 \Gamma}{\pi} \int_0^1 {\rm d} v  \int_v^1
\frac{ {\rm d}x}{\sqrt{1 -x^2}} \frac{ 2 x} {\left( \Gamma + 2 x\right)^2}
\nonumber \\
& =   & t\,\times \frac{4\Gamma}{\pi} \int_{0}^1 \frac{ {\rm d}x}{\sqrt{1 -x^2}}\left(  \frac{ 2 x} {\Gamma + 2 x}\right)^2 = C_1(\Gamma)\, t
\end{eqnarray*}
and recover the expected answer.

\section{Free fermions on ${\mathbb Z}^d$ with a source}
\label{sec:high-d}

An open quantum system of free fermions performing independent continuous time quantum walks and coupled to a localized source can be studied  on an arbitrary graph. Explicit calculations are
possible for  lattices admitting sufficiently simple lattice Green functions.
Thus, after describing the general formalism,  we focus on the square lattice.

\subsection{General Formalism}

In $d$ dimensions, we again  solve equation~(\ref{S-dual}) by  performing  Laplace and Fourier
transforms. We first define
\begin{subequations}
\begin{align}
\label{Sn-Lap-d}
\widehat{S}_{\bf n}(s) &= \int_0^\infty ds\,e^{-st} S_{\bf n}(t)\\
\label{Ssq-d}
S(s, {\bf q}) &= \sum_{{\bf n}}  \widehat{S}_{\bf n}(s)\,e^{- \ii {\bf
    q}\cdot {\bf n}}
\end{align}
\end{subequations}
where ${\bf n} = (n_1,\ldots,n_d)$, ${\bf q} = (q_1,\ldots,q_d)$
and ${\bf q}\cdot {\bf n}  =  q_1n_1   + \ldots+ q_d n_d $.
To avoid cluttered notation, we  write 
\begin{equation*}
\begin{split}
           \sum_{{\bf n}}       & =
                   \sum_{n_1=-\infty}^\infty \cdots
                   \sum_{n_d=-\infty}^\infty \\
          \label{dq}
\int d {\bf q} &=  \int_0^{2\pi}  \frac{dq_1}{2\pi}\, \cdots
\int_0^{2\pi}\frac{dq_d}{2\pi}       
\end{split}
\end{equation*}
 We also define 
\begin{subequations}
\begin{align}
\label{Psi-def}
\Psi_d(s, {\bf q})   & = \left[s - 2 \ii \sum_{a=1}^d\cos
  q_a\right]^{-1}\\
\label{Phi-def}
\Phi_d(s)              &= \left[\int d {\bf q}\,\Psi_d(s, {\bf q})
  \right]^{-1}
\end{align}
\end{subequations}
The function  $\Phi_d(s)$  is given through a $d-$fold integral which can be 
reduced  to a single integral involving a Bessel function. Indeed, writing $\Psi_d(s)$ as 
\begin{equation}
\label{ident}
\Psi_d(s, {\bf q}) = \int_0^\infty du\,\exp\!\left[-us+ 2u \ii
  \sum_{a=1}^d\cos q_a\right]
\end{equation}
and plugging \eqref{ident} into the definition of $\Phi_d(s)$ we
get 
\begin{equation*}
\frac{1}{\Phi_d(s)} = \int_0^\infty du\,e^{-us}\left[\int_0^{2\pi}
  \frac{dq}{2\pi}\, e^{2u \ii \cos q}\right]^d
\end{equation*}
Recalling an integral representation of the Bessel function 
\begin{equation}
\label{Bessel}
J_0(z) = \int_0^{2\pi} \frac{dq}{2\pi}\, e^{2z \ii \cos q}
\end{equation}
we express $\Phi_d(s)$ through a single integral:
\begin{equation}
\label{Phi-Bessel}
\Phi_d(s) = \left\{\int_0^\infty
du\,e^{-us}\left[J_0(2u)\right]^d\right\}^{-1}
\end{equation}

Applying  the Laplace-Fourier transform to the dual problem we arrive at
\begin{equation}
\label{SS0:d}
S(s, {\bf q}) = \left[1 -  \Gamma \widehat{S}_{\bf 0}(s)\right]
\Psi_d(s, {\bf q}) 
\end{equation}
From  the definition \eqref{Ssq-d}, we obtain 
\begin{equation}
\label{S0:d-int}
\widehat{S}_{\bf 0}(s) = \int d {\bf q}\,\,S(s, {\bf q})
\end{equation}
Using \eqref{SS0:d} and \eqref{S0:d-int} we fix the value of $\widehat{S}_{\bf 0}(s)$,
\begin{equation}
\label{S0:d}
\widehat{S}_{\bf 0}(s) = \left[\Gamma + \Phi_d(s)\right]^{-1} \, , 
\end{equation}
with $\Phi_d(s)$ defined in \eqref{Phi-def}. Therefore \eqref{SS0:d} becomes 
\begin{equation}
\label{S-d-exact}
S(s, {\bf q}) = \frac{\Phi_d(s)}{\Gamma+\Phi_d(s)}\, \Psi_d(s, {\bf
  q}) 
\end{equation}

The general relations \eqref{final} and \eqref{ampl} are valid in
arbitrary dimension; thus the growth rate is formally given by
\begin{equation}
  \label{Cind}
  C_d(\Gamma) = 2 \Gamma \left(1 -  2 \Gamma I \right) 
\end{equation}
 with 
\begin{equation}
\label{I-d}
I = \int_0^\infty {\rm d}t\, |S_{\bf 0}(t)|^2
\end{equation}
This integral
can  be expressed through the Laplace transform
$\widehat{S}_{\bf 0}(s)$, given in \eqref{S0:d},
via the Parseval-Plancherel identity \eqref{PP}. 
Hence, to  have more manageable results, one needs to
 compute $\Phi_d(s)$ explicitly 
to obtain an integral representation comparable to \eqref{int}. The `euclidean' versions of
integrals \eqref{Psi-def}--\eqref{Phi-def} arise in numerous
problems and are known as lattice Green functions (LGF). The LGF are reviewed
in Appendix~\ref{app:LGF} where we present explicit results for $\mathbb{Z}^d$ with $d=2$ and $d=3$, and briefly discuss other lattices. The LGF  is rather complicated even for the cubic lattice;
therefore,  in the following we focus on the square lattice.

\subsection{Growth rate for the process on square lattice}
\label{sec:2d}

For the square lattice, the integral in~\eqref{Phi-Bessel} can be
expressed through the complete elliptic integral of the first kind. One gets (see \cite{Hughes}  and  Appendix~\ref{app:LGF})
\begin{equation}
\label{Phi-2}
\Phi_2(s) = \frac{\pi s}{2K(4\ii/s)}
\end{equation} 
leading to
\begin{equation}
\label{S0:2}
\widehat{S}_{\bf 0}(s) = \left[\Gamma + \frac{\pi
    s}{2K(4\ii/s)}\right]^{-1}
\end{equation}
Using again the Parseval-Plancherel identity \eqref{PP}, we can write
\begin{equation}
\label{I-d2}
I = \lim_{\epsilon \to 0^+} \int_0^\infty dt\, |S_{\bf 0}(t)|^2
= \lim_{\epsilon \to 0^+} \int_{-\infty}^\infty  \frac{ {\rm d}y}{2 \pi}\,
\big|\widehat{S}_{\bf 0}\big(\tfrac{1}{2} \epsilon + \ii y\big)\big|^2
\end{equation}
The last expression  can be viewed as a complex integral along
vertical contour, infinitely close to the imaginary axis.
The function $K(z)$ is an even function, analytic in the complex plane
with branch cuts along the real axis from $-\infty$ to $-1$ and from
$1$ to $+\infty$. In the vicinity of the cuts, for $z = s \pm \ii \epsilon$
with $\Re(s) \ge 1$ and $\epsilon \to 0^+$, we have the following relation 
\cite{Abram,Byrd} 
\begin{equation}
\label{Kx:CUT}
K(z) = \frac{1}{s} \left(K\left(\frac{1}{s}\right) \pm \ii
K\left(\sqrt{ 1 - \frac{1}{s^2}}\right) \right)
\end{equation}
(see e.g., formula 162.02, page 39 in  \cite{Byrd}).
We  cut  the integral appearing in \eqref{I-d2} into four parts:
$ y \in ]-\infty, -4], [-4,0], [0,4]$ and $[4, \infty[$. Inside
the middle ranges, the function is analytic. In  the
boundary integrals, we   use  \eqref{Kx:CUT}\footnote{The correct sign is selected depending  
on which side  of the cut  the integrand lies.}. This leads to 
\begin{eqnarray}
  I = \int_{4}^{\infty} 
  \frac{ {\rm d}y}{2 \pi}\,
  \left| \Gamma + \frac{\ii \pi y}{2 K(4/y)} \right|^{-2}
  + \int_{0}^{4}  \frac{ {\rm d}y}{2 \pi}\,
 \left| \Gamma + \frac{2 \ii \pi}{ K(y/4) + \ii K(\sqrt{1 -y^2/16})} \right|^{-2} \nonumber  \\
 +\int_{-4}^{0}  \frac{ {\rm d}y}{2 \pi}\,\left|\Gamma - \frac{2 \ii \pi}{ K(y/4) - \ii K(\sqrt{1 -y^2/16})} \right|^{-2} +  \int_{-\infty}^{-4} \frac{ {\rm d}y}{2 \pi}\,
  \left| \Gamma + \frac{\ii \pi y}{2 K(4/y)} \right|^{-2}
\end{eqnarray}       
This integral representation leads to  the announced results \eqref{2d-main}, 
\eqref{I1} and  \eqref{I2}. 

\begin{figure}[t]
	\centering
	\includegraphics[width=0.6 \textwidth]{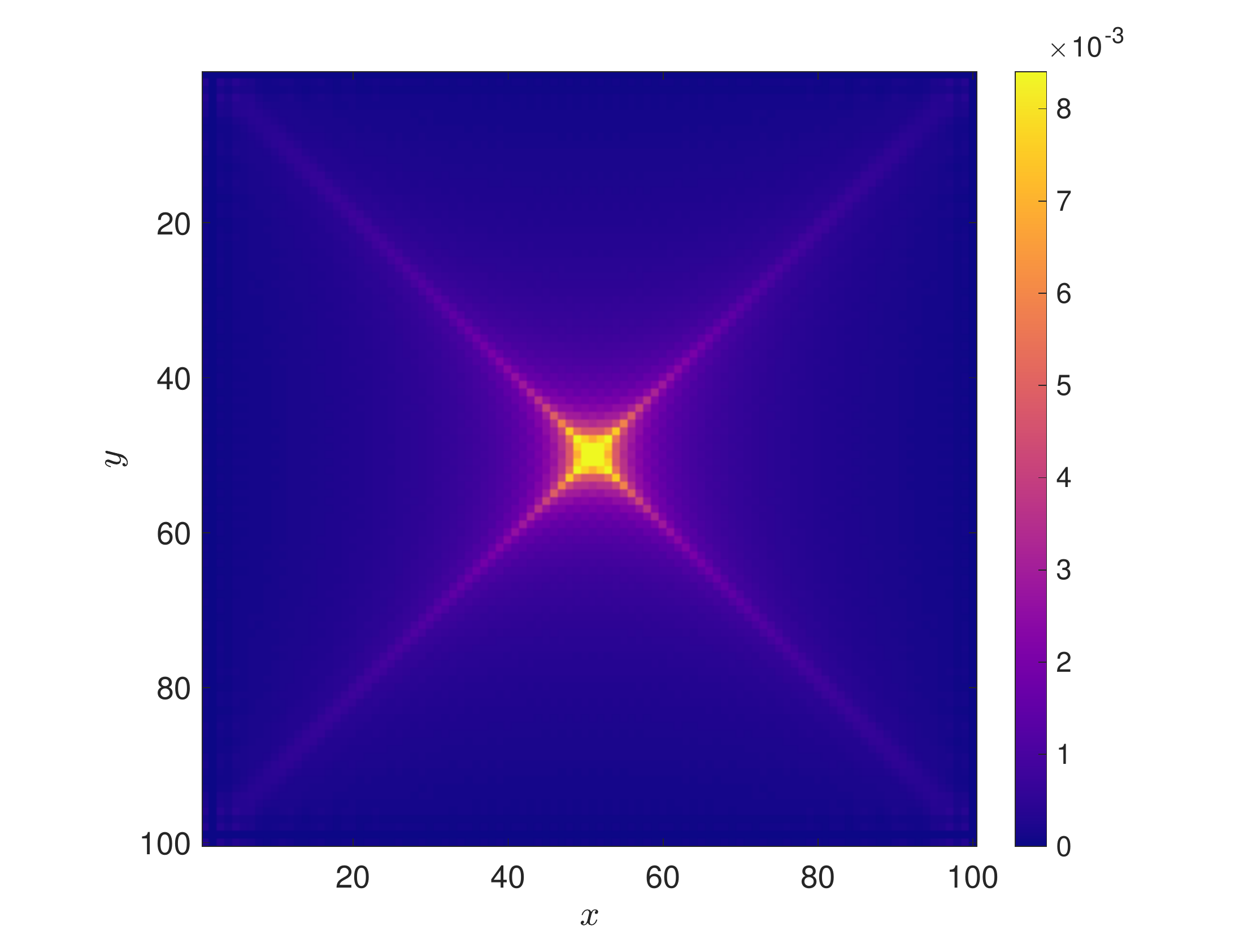}
	\caption{Density profile on a $100 \times 100$ square lattice at time $t=25$ for $\Gamma=0.1$.}
	\label{fig:density2D}
\end{figure}

To derive the  behavior for  $\Gamma\to\infty$, we start from 
 the asymptotic
\begin{equation}
\label{K-asymp}
K(x) = \frac{\pi}{2}\left[1+\frac{x^2}{4}+\ldots\right]
\end{equation}
Keeping the leading term in \eqref{K-asymp} we recast $I_2(\Gamma)$
into
\begin{eqnarray}
\label{I2-asymp}
\int_0^1 \frac{dx}{\Gamma^2 x^2+4^2} &=&
\frac{1}{4\Gamma}\,\tan^{-1}\frac{\Gamma}{4}
 = \frac{\pi}{8\Gamma} - \frac{1}{\Gamma^2} +
\frac{16}{3\Gamma^4} +O\big(\Gamma^{-6}\big)
\end{eqnarray}
Subtracting 
\eqref{I2-asymp} from \eqref{I2}  yields
\begin{eqnarray*}
I_2-\int_0^1 \frac{dx}{\Gamma^2 x^2+4^2} &=& \int_0^1
\frac{dx}{\Gamma^2 x^2+4^2}\,
\frac{4^2-[2\pi/\text{K}(x^2)]^2}{\Gamma^2 x^2+
  [2\pi/K(x)]^2}\\ &\simeq& 8 \int_0^1 dx\,\frac{x^2}{(\Gamma^2
  x^2+4^2)^2} \simeq  8 \int_0^\infty dx\,\frac{x^2}{(\Gamma^2
  x^2+4^2)^2} = \frac{\pi}{2\Gamma^3}
\end{eqnarray*}
Combining this with \eqref{I2-asymp} and 
 keeping terms up to $O(\Gamma^{-3})$, we obtain 
\begin{equation}
\label{I2-3}
I_2(\Gamma) = \frac{\pi}{8\Gamma} - \frac{1}{\Gamma^2} +
\frac{\pi}{2\Gamma^3}
\end{equation}
 Using  \eqref{I1} we
obtain a similar expansion  
\begin{equation}
\label{I1-3}
I_1(\Gamma) =  \frac{1}{\Gamma^2} - \frac{4\pi}{\Gamma^3}\int_0^1
dx\,\frac{K(x')}{K^2(x)+K^2(x')}
\end{equation}
Combining  \eqref{I2-3} and \eqref{I1-3} we arrive at
\begin{equation}
C_2(\Gamma) = \frac{B_2}{\Gamma}+O\big(\Gamma^{-2}\big)
\end{equation}
with 
\begin{equation}
B_2 = 64\int_0^1 dx\,\frac{K(x')}{K^2(x)+K^2(x')} - 8 
\end{equation}
 A numerical estimate shows
\begin{equation}
\label{ident}
\int_0^1 dx\,\frac{K(x')}{K^2(x)+K^2(x')} = \frac{1}{4}
\end{equation}
with very high precision. This  is probably an identity which would be interesting to prove. If \eqref{ident} is true, we get $B_2=8$,  leading to the announced result \eqref{large-2d}. For completeness, we show the $2D$ density profile in Fig.~\ref{fig:density2D}. Like in one dimension it has a pronounced peak near the injection site. Moreover, in contrast to the analogous classical problem, it is very anisotropic. It clearly displays interference of particles along high symmetry lines in the lattice. In principle the 1D analyses from section~\ref{sec:density} can be extended to 2D but analytic extraction of the asymptotics is more challenging and we leave it as an open problem. 

\section{Discussion}
\label{disc}

Theoretical studies of quantum gases with sinks, sources, and other
dissipative mechanisms have been mostly carried out using
one-dimensional models. Attractive features of  one-dimensional systems
include integrability of some toy models, and realizability in
experiments with cold atoms. In higher dimensions,  approximations and
simulations are required. For the simple quantum process  that we
studied, an open system of free fermions with injection from a
reservoir, we showed how to calculate the particle growth rate for
simple lattices in any dimension.  Analytical  formulae have been derived 
for the one-dimensional lattice and for the square lattice. With more
effort, explicit results for the particle growth rate could be 
derived for the triangular lattice, the cubic lattice, and perhaps
even for the body-centered hyper-cubic lattices as for such lattices
the lattice Green function can be expressed through hypergeometric
functions. More detailed characteristics can also be studied
analytically, e.g., we have calculated  the density  profile in one dimension. 
While the latter depend on the exact dispersion relations, linear growth of the average particle number is to be expected on any lattice for free fermions. The linear growth is simply induced by the fact that particles spread ballistically on any local translational invariant lattice Hamiltonian. 

In addition to the average total number of injected particles, one could 
explore  higher cumulants and  study the entire
distribution function $P(N,t)$, or  the 
full counting statistics. The large deviation (LD) formalism developed
for classical systems \cite{Ellis,D07,Var,HT} has been recently
applied to simple quantum systems such as single and two-qubit systems
\cite{Igor10,Budini,LLP11,Igor12}. Large open systems of
non-interacting fermions are also tractable in stationary
one-dimensional settings, for example  non-equilibrium steady states of
boundary driven systems have been established \cite{KP09,Z10a}, and
the large-deviation function has been subsequently calculated
\cite{Z14}. The stationarity is crucial in the procedure
\cite{Prosen08} employed in the calculation of the large-deviation
function \cite{Z14}. For other applications of the LD formalism to
boundary-driven non-interacting quantum chains, see
Refs.~\cite{SK13,Z14b,Prosen14,Swingle16,Cecile17,Igor17,Igor18}.

Our system is manifestly non-stationary as the number of particles
increases with time. In some evolving free fermion systems, 
fluctuations have been probed analytically;  for instance, in a 
system of lattice fermions starting from the domain wall initial
condition, the number of fermions in the initially empty half-line
grows linearly in time, while the variance increases logarithmically
\cite{Hungary08}. Recently, the full statistics of the current in this
problem has been calculated using powerful  techniques inspired by
previous studies of classical stochastic systems \cite{Japan19}.  
We believe that system could be also explored using
integrable probability methods.

We totally ignored interactions. A natural generalization of the
present model is the XXX chain with all spins originally
pointing down in which the spin at the origin can be flipped
upwards with a certain rate.  The equations for the correlators in
this model are hierarchical rather than recurrent  and the
factorization property that played a key role in the present study
also does not hold. Various matrix representation techniques
have been applied to open interacting finite spin chains coupled to reservoirs
\cite{ProsenXXZ,ProsenXXZ2,ProsenCarlos} and it would be interesting
to extend these results to the infinite chain case (see
\cite{ProsenReview} for a review and references).

There are many other open quantum systems with local sinks or source
which can be potentially treated. The most obvious extension is to
bosons, see \cite{Spohn10,Dries18,KMS} for work in this
direction. For example, in contrast to fermions, a phase transition was reported in~\cite{Dries18} in the case of weakly interation bosons with a local sink.

\vskip 0.666cm
\noindent 
{\bf Acknowledgments.} We are grateful to Jean-Marc Luck, Michel Bauer and 
Toma\v{z} Prosen for stimulating discussions, and to S. Mallick for
a careful reading of the manuscript.  PLK thanks the Institut de
Physique Th\'eorique for hospitality. DS acknowledges support from the FWO as post-doctoral fellow of the Research Foundation -- Flanders.

\appendix

\section{Single particle case}
\label{App:density-dual-1d-singlep}

In this appendix we determine the density profile in the dual problem starting with a 
single particle at the origin; we consider the most symmetric setting when the sink is also  at the origin.  The inverse Laplace transform of \eqref{S0} admits an
integral representation \cite{PBM}
\begin{equation}
\label{S0-sol}
S_0(t) = e^{-\Gamma t} - 2\int_0^t du\,J_1(2u)\,
e^{-\Gamma\sqrt{t^2-u^2}}
\end{equation}
In the long time limit, the integral term dominates. Computing its
asymptotic behavior we find
\begin{equation}
S_0(t)\simeq -\frac{2 J_1(2t)}{\Gamma^2 t}\simeq -
\frac{2}{\Gamma^2}\,\frac{1}{\sqrt{\pi
    t^3}}\,\cos\!\left(2t-\frac{3\pi}{4}\right)
\end{equation}
The density at the origin 
\begin{equation}
\label{N0:asymp}
N_0(t) = |S_0(t)|^2 \simeq \frac{4}{\pi \Gamma^4
  t^3}\left[\cos\!\left(2t-\frac{3\pi}{4}\right)\right]^2
\end{equation}
oscillates indefinitely reflecting quantum interference. Further, the
density at the origin vanishes infinitely often: $N_0(t_k)=0$ at times
$0<t_1<t_2<t_3<\ldots$ significantly depending on $\Gamma$ for
$k=O(1)$ and behaving universally, $t_k\simeq \tfrac{\pi}{8}+
\tfrac{\pi}{2}k$, for $k\gg 1$. 

After averaging \eqref{N0:asymp} over oscillations we arrive at 
\begin{equation}
\label{N0:asymp-av}
N_0(t)  \simeq \frac{2}{\pi \Gamma^4 t^3}
\end{equation}

\begin{figure}
\centering \includegraphics[width=7.77cm]{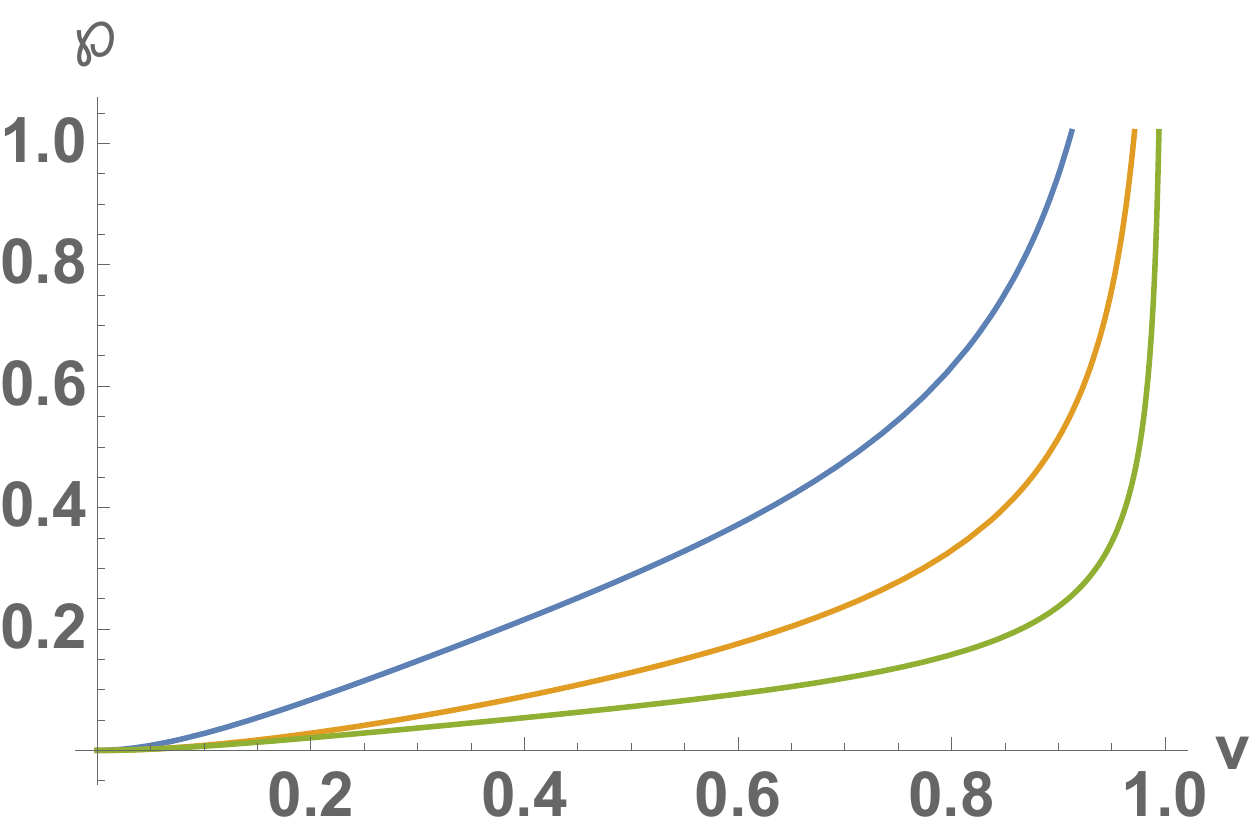}
\caption{The scaled density profiles for the dual problem
  in which a particle is removed from site 0  by a trap of  strength $\Gamma$. 
  Analytically, $\wp(v; \Gamma)$ is defined via \eqref{Sn-scaled}. The
  curves (top to bottom) correspond to 
  $\Gamma=1,2, 4$.}
\label{Fig:density-3}
\end{figure}

When $n>0$, the inverse Laplace transform of \eqref{Sn-Lap} admits an
integral representation
\begin{eqnarray}
\label{Sn-sol}
S_n(t) &=& J_n(2t) \nonumber\\ &-& \Gamma\int_0^t du\,e^{-\Gamma
  u}\left(\frac{t-u}{t+u}\right)^\frac{n}{2}J_n\!\left[2\sqrt{t^2-u^2}\right]
\end{eqnarray}
In the scaling limit
\begin{equation}
\label{scaling}
|n|\to\infty, \quad t\to\infty,  \quad v=\frac{n}{2t}=\text{finite}
\end{equation}
we have
\begin{equation*}
\left(\frac{t-u}{t+u}\right)^\frac{n}{2}\simeq e^{-2v u}
\end{equation*}
and therefore \eqref{Sn-sol} yields
\begin{equation}
\label{Sn-scaling}
S_n (t) \simeq \frac{2v}{\Gamma+2v}J_n(2t)
\end{equation}
Using the asymptotic of $J_n(2t)$ in the scaling limit \eqref{scaling}
we find that $N_n=|S_n|^2=0$ when $|v|>1$ and
\begin{subequations}
\begin{equation}
\label{Sn-scaling-2}
N_n(t) \simeq (2\pi t)^{-1}\, \wp(v; \Gamma)
\end{equation}
with the scaled density given by
\begin{equation}
\label{Sn-scaled}
\wp(v; \Gamma) =\left(\frac{2|v|}{\Gamma+2|v|}\right)^2
\frac{1}{\sqrt{1-v^2}}
\end{equation}
\end{subequations}
when $|v|<1$. Equation \eqref{Sn-scaled} has been obtained by
averaging over oscillations.  The scaled density profiles \eqref{Sn-scaled}
are shown in Fig.~\ref{Fig:density-3} for three values of the strength $\Gamma$
of the trap.  As a useful consistency check, we recover
 the value \eqref{>2} of the asymptotic  survival probability 
\begin{eqnarray*}
\Pi_\infty  =  \sum_{n=-\infty}^\infty |S_n(\infty)|^2
= \frac{1}{\pi} \int_{-1}^1 \frac{dv}{\sqrt{1-v^2}}
\left(\frac{2|v|}{\Gamma+2|v|}\right)^2 
\end{eqnarray*} 

\section{Lattice Green functions}
\label{app:LGF}

One can relate the Euclidean version of the integrals \eqref{Psi-def}--\eqref{Phi-def} to formulae arising  in the context of lattice random walks \cite{Hughes}. Let $P_k$ be the probability that a random walker starting at the origin returns back after $k$ steps. The generating function $P(z) = \sum_{k\geq 0} P_k z^k$, known as the lattice Green function (LGF), admits an
integral representation 
\begin{equation}
\label{LGF}
P_d(z) = \int \frac{d{\bf q}}{1-z\Lambda({\bf q})}
\end{equation}
with 
\begin{equation}
\Lambda({\bf q}) = \Lambda_d({\bf q}) \equiv
\frac{1}{d}\sum_{a=1}^d\cos q_a
\end{equation}
for the hyper-cubic lattice. The general formula \eqref{LGF} is valid
for many other lattices, the specificity of the lattice is reflected
in the function $\Lambda({\bf q})$. For instance, 
\begin{equation*}
\begin{split}
\Lambda_\triangle({\bf q})   &= \frac{\cos q_1 + \cos q_2 +  \cos(q_1+q_2)}{3}\\ 
\Lambda_\text{bcc}({\bf q}) &= \cos q_1 \cos q_2 \cos q_3 \\ 
\Lambda_\text{fcc}({\bf q}) &= \frac{\cos q_1 \cos q_2 +
  \cos q_2 \cos q_3 + \cos q_1 \cos q_3}{3}
\end{split}
\end{equation*}
for the triangular lattice, the body-centered cubic (bcc) lattice and the face-centered cubic (fcc) lattice. The LGFs for many
low-dimensional lattices are derived in a book by Hughes \cite{Hughes}; for more recent results, see
\cite{Glasser,Joyce,Joyce03,Guttmann} and references therein.  Here we outline some remarkable special cases. 

\vskip 0.3cm

In two dimensions, for  the square lattice, the LGF admits a remarkably simple expression
through the complete elliptic integral (or as an  hypergeometric function)
\begin{equation}
\label{LGF:square}
P_2(z) = \frac{2}{\pi}\,K(z)
 = {}_2F_1\!\left(\tfrac{1}{2}, \tfrac{1}{2}; 1; z^2\right)
\end{equation}
We have used this expression to  derive \eqref{2d-main} and
\eqref{I1}--\eqref{I2}.  For  a two-dimensional triangular lattice, the LGF has a more complicated form
\begin{equation}
\label{LGF:triangular}
P_\triangle(z) = \frac{6}{\pi z \sqrt{(\zeta_+-1)(\zeta_- + 1)}}\,K(k)
\end{equation}
with  $k = \sqrt{\frac{2(\zeta_+ - \zeta_-)}{c}}$ and
 $ \zeta_\pm  = \frac{3}{z}+1 \pm \sqrt{3+\frac{6}{z}}\, .$

\vskip 0.3cm

In three dimensions, for the body-centered cubic lattice,
a neat expression for the LGF has
been found long ago \cite{Montroll}
\begin{equation}
\label{LGF:bcc}
P_\text{bcc}(z) = \left[\frac{2}{\pi}\,K(k_2)\right]^2, \quad 2 k_2^2
= 1 - \sqrt{1-z^2}
\end{equation}
For the three-dimensional cubic lattice, the LGF reads \cite{Joyce}
\begin{equation}
\label{LGF:cubic}
P_3(z) =
\frac{1-9\xi^4}{(1-\xi)^3(1+3\xi)}\left[\frac{2}{\pi}\,K(k_1)\right]^2
\end{equation}
with  $k_1^2 = \frac{16\xi^3}{(1-\xi)^3(1+3\xi)}$ and $ \xi^2   =
\frac{1-\sqrt{1-z^2/9}}{1+\sqrt{1-z^2}}\, .$

\vskip 0.3cm

In four and higher dimensions, the LGF
of  the  body-centered hyper-cubic lattices  can be expressed
\cite{Guttmann} through the hypergeometric function in arbitrary
dimension 
\begin{equation}
\label{LGF:hyper-bcc}
P_\text{d-bcc}(z) = {}_dF_{d-1}1\!\left(\tfrac{1}{2},
\ldots,\tfrac{1}{2}; 1,\ldots,1; z^2\right)
\end{equation}
In contrast, the calculations  of the LGF of 
the hyper-cubic lattices require  rather
abstract mathematics such as Calabi-Yau ordinary differential
equations \cite{Glasser,Joyce03,Guttmann}.

\end{document}